# EEG-based listened-language classification

Isaac Ariza [a], Lorenzo J. Tardón [b,*], Ana M. Barbancho [b,*], Isabel Barbancho [b,*]

[a] ATIC Research Group, E.T.S.I. Telecomunicación, Universidad de Málaga, Málaga, Spain
[b] ATIC Research Group, ITIS Software, E.T.S.I. Telecomunicación, Universidad de Málaga, Málaga, Spain

A B S T R A C T

From an early age, individuals are continuously exposed to other languages beyond their native tongue; however, the brain's response to these auditory stimuli remains unclear. To investigate this, an experiment was designed to record electroencephalography (EEG) signals from subjects listening to sentences in five different languages, and a specific database was built to enable performing classification tests to distinguish between different languages, and varying levels of language comprehension. By analysing the energy difference between the EEG channels to characterize these signals, different classification tests were conducted using bidirectional Long Short-Term Memory (bi-LSTM), Long Short-Term Memory (LSTM) and Gated Recurrent Unit (GRU) neural networks.
The main objective is the analysis of the brain's response in two different scenarios: when the subject listens to sentences in different languages, and when the subject understands or misunderstands the meaning of a sentence. In the multi-class classification involving sentences in five different languages, the accuracy attained is 36.37 %. However, in the multi-class classification between 'understood'/'understood part of the meaning'/'didn't understand', the accuracy attained reaches 81.36 %. The results obtained for binary classification tests of understand native language or foreign language is 89.09 %. The bi-LSTM neural network achieved the overall best performance.
These results demonstrate that the analysis of the EEG signals alone can give information regarding a person's language comprehension level, and can be used for monitoring the learning curve of a new language or to assess comprehension in patients with conditions such as aphasia.

## 1. Introduction

In today's globalized world, proficiency in two or more languages offers significant social, cultural, academic, and professional benefits, which gives rise to the study of its impact on the social development of individuals, as noted in Movsum (2020). The ability to communicate in multiple languages is known as bilingualism or multilingualism, and numerous studies have been conducted to determine how it affects brain function, especially in children (Ali, 2023).

Li, Zhang, Ding, Zhou, and Yu (2019) use functional near-infrared spectroscopy (fNIRS) technique to explore the effect of bilingualism on cognitive performance in young children, and analyse electroencephalography (EEG) signals to investigate the mechanism of bilingualism (Li & Song, 2023). Liu, Xing, Huang, Schwieter, and Liu (2023) investigate how bilinguals control language switching. In a study of EEG signals, N400 components of monolingual and bilingual subjects listening to a story were compared (Momenian, Vaghefi, Sadeghi, Momtazi, & Meyer, 2024). In monolingual subjects, the N400 varies according tothe current and previous word heard, and in bilingual subjects only the previously heard word is reflected, leading to a delay in the processing of story comprehension. Other studies have focused on how bilingualism affects the neural mechanisms of attention (Chung-Fat-Yim, Bobb, Hoshino, & Marian, 2023) or whether it improves cognitive performance (Swenson, 2023).

The most widely used techniques for understanding brain reactions, are fNIRS and EEG mentioned above. EEG, in particular, allows obtaining real-time measurements of EEG signals produced by neuronal activity of the brain. The existing literature on EEG signal processing is extensive and pursues different objectives such as artefact removal (Agounad, Tarahi, Moufassih, Hamou, & Mazid, 2025; Sharma & Meena, 2024), feature extraction (Pooja, Pahuja, & Veer, 2022; Singh & Krishnan, 2023) or noise reduction (Pant & Kumar, 2024), etc. By applying digital signal processing (Sanei & Chambers, 2008) and machine learning algorithms (Tripathy & Pachori, 2024) such as neural networks (Pratiwi, Wibawa, & Purnomo, 2021; Rabbani & Islam, 2024; Xu & Xia, 2023; Zhang, Wang, Zhang, & Chen, 2019), it is possible to obtain information about the brain's reaction to different stimuli, such as listening to voices or music (Ariza, Barbancho, Tardón, & Barbancho, 2023), or the location of the sources of these responses.

* Corresponding authors.
*E-mail addresses:* iariza@ic.uma.es (I. Ariza), lorenzo@ic.uma.es (L.J. Tardón), abp@ic.uma.es (A.M. Barbancho), ibp@ic.uma.es (I. Barbancho).

Machine learning techniques, including deep learning techniques, are being applied the most diverse fields, as (Agounad et al., 2025; Sharma & Meena, 2024), feature extraction (Pooja et al., 2022; Singh & Krishnan, 2023), noise reduction (Pant & Kumar, 2024), image classification (Chen, Sun, Li, Zheng, & Chen, 2025; Zheng, Liang, Zhao, & Deng, 2024), aviation safety (Guo et al., 2024; Lin et al., 2023), resource management (Zhu, Li, Chen, Zhou, & Deng, 2024), etc. Within our context, neural networks, and, specifically, Long Short-Term Memory (LSTM) networks, are widely used for medical applications and biosignal analysis. Zhang, Yao, Yang, Zheng, and Liu (2022) use LSTM neural networks to predict heart motion to improve the accuracy and safety of robot-assisted cardiac surgery. In contrast, Yang et al. (2023) employ a Gated Recurrent Unit (GRU) network structure for the same purpose and obtain better results. Hybrid models combing Convolutional Neural Networks (CNN) and LSTM networks are used to detect Autism Spectrum Disorder (Lakhan, Mohammed, Abdulkareem, Hamouda, & Alyahya, 2023) or to classify histopathological images for breast cancer detection (Srikantamurthy, Rallabandi, Dudekula, Natarajan, & Park, 2023). GRU and bidirectional LSTM architectures are also used for emotion recognition (Maliha, Lopa, & Chowdhury, 2024).

Many diverse applications can emerge from the analysis of EEG signals: brain computer interface (BCI) based on EEG (Pfeffer, Ling, & Wong, 2024) with different purposes: play the game of "whack a mole" (Jahangiri, Achanccaray, & Sepulveda, 2019) or control a robotic arm (Zhou et al., 2023), detection of sleep disorder using a convolutional neural network using EEG signals as input (Sudhakar et al., 2021), or emotion recognition (Moschona, 2020; Rahman et al., 2021), among others.

Some research groups have performed EEG signal analysis in the context of language learning. In Syed and Wang (2018), a robotic assistant monitors EEG signals from students during Chinese lessons to assess their concentration and interest level. With this, if the concentration level of the student is low, words can be repeated. Alimardani, Duret, Jouen, and Hiraki (2022) investigate the brain responses of children during a French lesson in which they listen to a story told by a robot or directly on a computer screen; and Nakov and Alimardani (2022) analyse EEG signals of children learning a second language using machine learning techniques such as Support Vector Machine (SVM) or K-Nearest Neighbors (KNN) to measure learning performance. In relation to language learning, the EEG signal can also help determine whether the learner understands single words, as described in Schneegass, Kosch, Schmidt, and Hussmann (2019).

Studies on EEG signals related to language have also been conducted for other purposes, such as mapping the brain areas that control language using Convolutional Neural Networks (CNN) (Adhikari, Pham, Hall, Rotenberg, & Besio, 2022), improving speech recognition, including silent speech recognition, (Vorontsova et al., 2021), or examining the differences between subjects for whom the language is native and those for whom it is a second language (Sakthi, Tewfik, & Chandrasekaran, 2019). Also, EEG and magnetoencephalography (MEG) have been combined to get a better understanding of the process of language comprehension (Wang & Kuperberg, 2024). Dogan, Tuncer, Barua, and Acharya (2024) perform bilingual classifications using EEG signals from subjects reading and listening to content in English and Turkish.

Within this context, some gaps have been identified in the literature in this context:

- Studies on EEG-based language recognition typically involve only two different languages.
- There is a lack of literature on EEG-based classification for assessing the level of understanding of sentences in different languages.

In this framework of analysis of brain activity in relation to language and comprehension, the research presented in this manuscript is motivated by the need to enhance our knowledge on the brain's response to auditory stimuli of different languages, both familiar and unfamiliar, native and foreign languages, and examine the brain's reaction when the meaning of a sentence is either understood or misunderstood to, ultimately, enable the research for tools to aid patients in a state of minimal consciousness or with communication difficulties to find alternative connection means.

The focus of this article is on identifying or developing tools or schemes to classify the brain's response to different languages through the analysis of EEG signals in order to know if the subjects under test understand what they hear or not, and to know if it is their native language or a foreign one. Thus, this work contributes to the theoretical understanding of the brain within the context of cognitive processes associated with communication, and, specifically, speech. This fits in the BCI development framework for EEG signal analysis for speech communication (Jahangiri et al., 2019; Pfeffer et al., 2024). For this purpose, a specific experiment has been designed in which the subjects under test listen to sentences in different languages; the five languages selected for the experiment are Spanish, English, German, Italian and Korean. Brain responses reflected in the EEG data are later analysed and processed to find their relation to language and understanding. Note that in the experiment, their native language and other languages are used; then the data obtained are used to carry out binary classification and multi-class classification tasks.

The novelty of this study lies primarily in the languages used, as they are very different from each other; also the parameters used to characterize each trial, and the use of different neural architectures to analyse and compare results are novel contributions in our context. The main difference from existing literature is the incorporation of five languages, both familiar and unfamiliar for the subjects; additionally, we get feedback on the level of understanding of each sentence. This allows us to know the brain's reaction when a subject does not understand a sentence. Another novel aspect is the ability to identify the language a subject is listening to by analysing their EEG signals. With this, different aspects have been analysed, such as the level of understanding of the language and whether it is the native language of the subject.

To characterize each trial, a new matrix of energy differences between EEG channels has been developed. In addition, the evolution of energy differences between consecutive trials is considered. Three different neural network architectures have been employed in this analysis: Long Short-Term Memory (LSTM), Bidirectional Long Short-Term Memory (bi-LSTM) and Gated Recurrent Unit (GRU). These architectures are used to compare the results obtained from each approach.

Binary classification and multi-class classification tasks have been carried out. In the multi-class classification involving sentences in five different languages, the accuracy achieved is 36.37 %. However, in the multi-class classification task distinguishing between 'understood'/'understood part of the meaning'/'didn't understand', the accuracy attained reaches 81.36 %. The results obtained for the binary classification tasks that differentiate between native language or foreign language the accuracy reaches 89.09 %.

This paper is organized as follows. In Section 2, the designed experiment, the dataset employed and the data acquisition methodology are described. Next, in Section 3, the proposed method is outlined, including the trial segmentation process for feature extraction. The features themselves, and the structure of the recurrent neural networks employed are described next. The results found after the experiments performed, along with a discussion on the results are presented in Section 4. Finally, some conclusions drawn from the work presented are exposed in Section 5.

## 2. Dataset employed

This section describes the dataset and data acquisition procedure employed in this study. First, in Section 2.1, the equipment used to capture the EEG signals, and the conditions under which the recordings were made are described. In Section 2.2, the specific language-related experiment designed for the research is explained. Finally, in Section 2.3, the characteristics of the dataset are presented.

**Table 1**
Dataset: characteristics of the sentences in different languages, and labels used to identify each language.

| Language | Label | # Of samples | Total duration |
|---|---|---|---|
| Spanish | SP | 30 | 2:20 |
| English | EN | 30 | 2:03 |
| Italian | IT | 30 | 2:30 |
| German | GE | 30 | 1:59 |
| Korean | KO | 30 | 1:46 |

### 2.1. EEG Capture system

A BrainVision actiCHAMP-PLUS system is used to record the EEG signals. This system has been employed in other works with a similar configuration (Ariza et al., 2023). It has a total of 160 channels, of which 64 are employed in this work. The sampling rate is 2500 Hz, and the electrodes are positioned according to the 64 channel actiCAP snap standard (BrainVision, 2024).

Channel FCz is used as reference channel, while FPz serves as ground channel. The electrodes of channels FT9 and FT10, near the eyes, are used to capture vertical (VEOG), and horizontal (HEOG) ocular movements, respectively. The channels used for eye movement are, consequently, discarded regarding EEG analysis; consequently, 61 EEG channels are finally used in this work. The impedance at all the electrodes is measured, making sure that their impedance is below 0 KΩ in all cases, and maintains those low levels throughout the entire experiment.

The EEG signals are segmented into trials of two different durations: 400 ms and 1 s, with a 50 % overlap. Mean value and linear trends per trial are removed at the pre-processing stage.

### 2.2. Experiment: listening to phrases in different languages

#### 2.2.1. Experiment design

The experiment designed for this research involves listening to a series of short sentences, about 5 second long, in five different languages: Spanish, English, Italian, German and Korean. Note that most works on language recognition use only two different languages; also, some of the languages selected are very different, with diverse root, from each other.

Each subject listens to 30 sentences per language, presented in groups of 5 sentences for each language. The order in which the group of 5 phrases of the different languages are listened to is randomized. To facilitate later identification of each sentence, a unique id is assigned to each sentence. The labels employed to mark each sample are shown in Table 1.

All sentences used in this experiment are listed in Appendix A. Note that colloquial phrases commonly used in everyday life were selected for this study (Tables A.1–A.5).

After listening to each sentence, the subject is asked to assess their level of understanding with one of three possible answers: understood the meaning, partially understood the meaning or did not understand the meaning of the sentence heard.

The software program E-prime 3 was used to present stimuli in this experiment (Psychology Software Tools, 2024). E-prime 3 is commonly used in this field to present different stimuli, either auditory or visual ones. It is also responsible for creating time stamps in the recording of EEG signals, which facilitate the synchronization and identification of trials in different languages, and for collecting the subjects' responses to the questions asked after listening to each sentence.

#### 2.2.2. Experimental setup arrangement

Due to the nature of EEG signals, which have amplitudes in the microvolt range, special care must be taken during signal acquisition. Therefore, the experiments are conducted in an isolated room to minimize noise and distractions. The experiment is conducted individually. The subject sits in a chair without armrests in front of a computer, listens to the sentences through speakers, and enters their responses using a keyboard. In Fig. 1, an illustration of the complete setup for data acquisition is shown.

Prior to the experiment, the subjects were instructed to relax and avoid movement in order to reduce artefacts in the recorded EEG signals.

#### 2.2.3. Subjects under test

This experiment involved 6 healthy people, 2 men and 4 women, with an average age of 22.33 years. The specific objective was to investigate the brain's to sentences listened in their native language or in a foreign one. It is important to note that the native language of all the subjects is Spanish, except for one of them who is from China.

Prior to the start of the experiment, participants completed a questionnaire assessing their level of language proficiency. They were asked to rate their understanding on a scale from 1 to 5, where 1 means no understanding and 5 native-level understanding. Additionally, they were asked whether they know any other languages. Table 2 shows the responses of the subjects who participated in this experiment.

### 2.3. Dataset characteristics

The dataset used this study comprises EEG signals recorded from six subjects during the experiment described in the Section 2.2. The data acquisition methodology was conducted according to the guidelines of Declaration of Helsinki, and approved by the *Comité de Ética de la Investigación Provincial de Málaga* on December 19, 2019, approval number: 2176-N-19. This database has been published, and it is available for use by the research community (Ariza, Barbancho, Tardón, & Barbancho, accessed 10 March 2025).

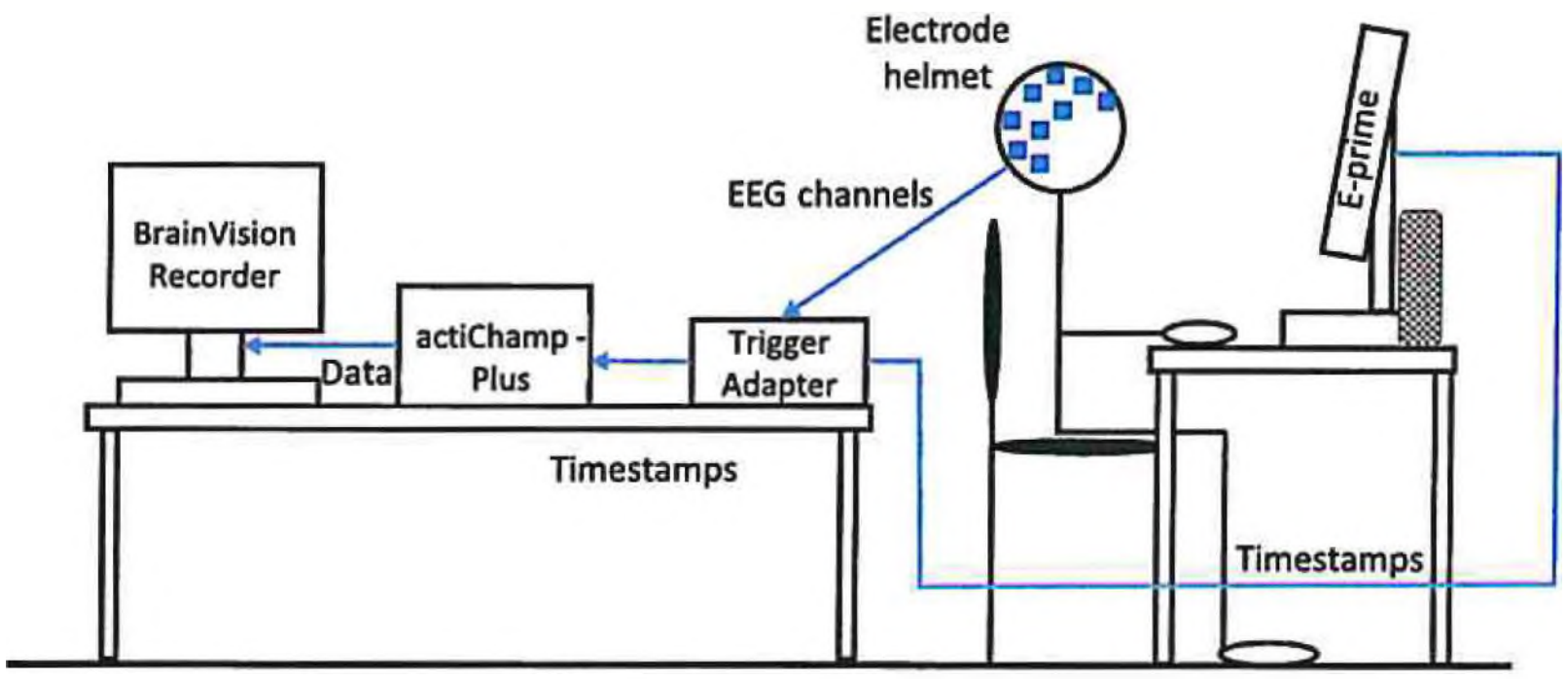

**Fig. 1.** Setup for data acquisition.

**Table 2**
Dataset: subjects.

| Subject | Spanish | English | Italian | German | Korean | Other language |
|---|---|---|---|---|---|---|
| 0009 | 4 | 4 | 1 | 1 | 1 | Chinese and Japanese |
| 0010 | 5 | 3 | 1 | 1 | 1 | French |
| 0013 | 5 | 3 | 1 | 1 | 1 | French |
| 0014 | 5 | 3 | 1 | 3 | 1 | – |
| 0015 | 5 | 4 | 1 | 1 | 1 | – |
| 0017 | 5 | 4 | 4 | 1 | 1 | French |

**Table 3**
Dataset: number of trails per language.

| Language | Label | Number of 400 ms trials | Number of 1 s trials |
|---|---|---|---|
| English | EN | 3715 | 1687 |
| German | GE | 3739 | 1699 |
| Italian | IT | 3821 | 1736 |
| Korean | KO | 3692 | 1676 |
| Spanish | SP | 3686 | 1665 |

The duration of the experiment is over 16 minutes, with a sampling rate of 2500 Hz, and hardware low-pass filtering with cut-off frequency of 690 Hz. Bias and long-term linear trends are removed. The samples were segmented into different types, each further divided into trials of two different sizes: 400 ms and 1 s of duration, with a 50 % overlap. These trial durations were chosen to capture the P300 event-related potential (ERP) generated after the presentation of the auditory stimuli. The P300 event-related potential is a positive-going potential, peaking at around 300 ms after stimulus. This component represents the moment at which a stimulus elicits a brain response (Miranda & Castet, 2014). Table 3 shows the number of each type of trial for each language and trial duration.

While the number of subjects is modest, the length of the experiment allows a large number of samples for the classification tasks to perform.

## 3. Analysis methodology

In this section, we describe the data preparation and processing steps conducted prior to the classification process, which imply the characterization of EEG brain activity by means of custom relative energy arrays, and their temporal evolution. Then, the recurrent neural networks selected for our concrete classification tasks and particular working conditions, together with their specific design parameters, will be described.

The segmented signals, with trial durations of 400 ms or 1 s, as described in Section 2.3, are processed to compute the energy difference matrix, which is used to characterize each trial (see Section 3.1). Finally, the architecture of the recurrent neural networks employed for signal classification is shown in Section 3.2. A diagram of the whole processing scheme is shown in Fig. 2. Details of each of the processing stages are given in the next subsections.

### 3.1. Energy difference matrix

In this study, we use EEG signals acquired with the recording system described in Section 2.1. These data encompass 61 time-domain signals, one for each channel. These signals are segmented into 400 ms or 1 s trials with a 50 % overlap. Each time-domain EEG signal is denoted as $eeg_{n,m}(t)$, representing the EEG signal of channel $m$ in trial $n$.

Brain activity will be considered in terms of relative measured energy across all the EEG channels, similarly to Ariza et al. (2023). To this end, the signals are first transformed to the frequency domain by applying the Fast Fourier Transform (FFT) (Mathworks help center - fft, 2024), resulting in $EEG_{n,m}(f)$, which represents the Fourier transform of the EEG signal of channel $m$ in trial $n$. The energy of channel $m$ in trial $n$, denoted as $e_{n,m}$ is obtained from $EEG_{n,m}(f)$, according to Parseval's theorem, as defined in Oppenheim, Willsky, and Hamid-Nawab (1997). Due to the very low amplitude of EEG signals, a logarithmic scale is employed to better represent the energy. Therefore, $E_{n,m}$, which represents the energy of the EEG signal of channel $m$ in trial $n$ expressed in decibels (dB) is calculated using the following formula (Ariza et al., 2023):

$$E_{n,m}(dB) = 10log_{10}(\int_{-\infty}^{\infty} \|EEG_{n,m}(f)^2\|df) \quad (1)$$

Once the energy for each channel, $E_{n,m}(dB)$ is obtained, we compute the matrix that characterizes each trial by calculating the relationships between energies in the different channels, in dB. This procedure results in the following $61 \times 61$ matrix (Ariza et al., 2023):

$$\mathbf{E(n)} = \begin{pmatrix} 0 & E_{n,1} - E_{n,2} & \cdots & E_{n,1} - E_{n,61} \\ E_{n,2} - E_{n,1} & 0 & \cdots & E_{n,2} - E_{n,61} \\ \vdots & \vdots & \ddots & \vdots \\ E_{n,61} - E_{n,1} & E_{n,61} - E_{n,2} & \cdots & 0 \end{pmatrix} \quad (2)$$

Additionally to $\mathbf{E(n)}$, the evolution of energy differences between consecutive trials is considered. To achieve this, the first derivative is observed, which is approximated by using the centred difference formula. Thus, $\mathbf{E}'(\mathbf{n})$, defined as:

$$\mathbf{E}'(\mathbf{n}) = \frac{\mathbf{E(n+1)} - \mathbf{E(n-1)}}{2\Delta} \quad (3)$$

with $\Delta = 1$, is calculated to explicitly consider the evolution of the relationship between the energies obtained from the measured EEG channels (Ariza et al., 2023).

### 3.2. Recurrent neural network

Recurrent neural networks (RNNs) are a type of neural network designed to capture temporal characteristics from sequences of data for characterization. Among RNNs, Long Short-Term Memory (LSTM) networks are a sub-type of RNNs equipped with specialized nodes that control the flow of information, and extract features for the classification of temporal sequences, as in Ariza et al. (2023), Ariza, Tardón, Barbancho, De-Torres, and Barbancho (2022), Pamungkas, Wibawa, and Rais (2022), Chakkamallisery et al. (2023). Gated Recurrent Unit (GRU) networks constitute an alternative solution to alleviate the complexity of LSTM units, while still being effective for the classification of temporal sequences, as evidenced in Yang et al. (2023) and Chowdary, Anitha, and Hemanth (2022).

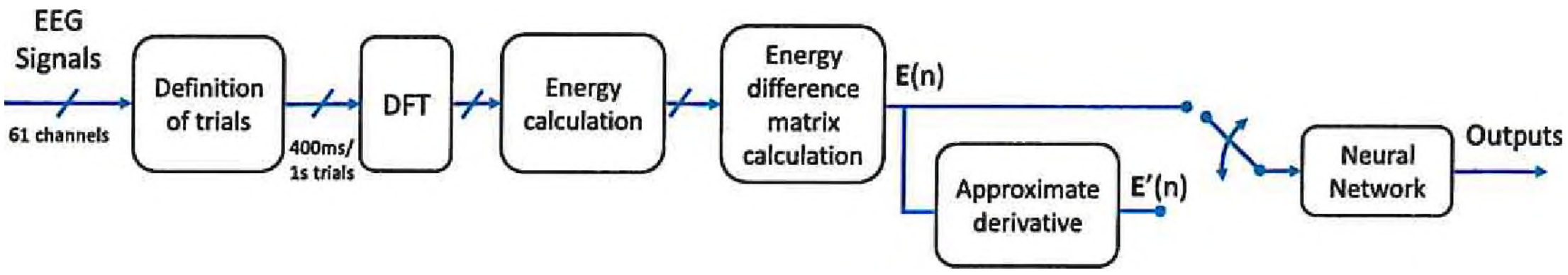

Fig. 2. General diagram of the EEG signal processing scheme.

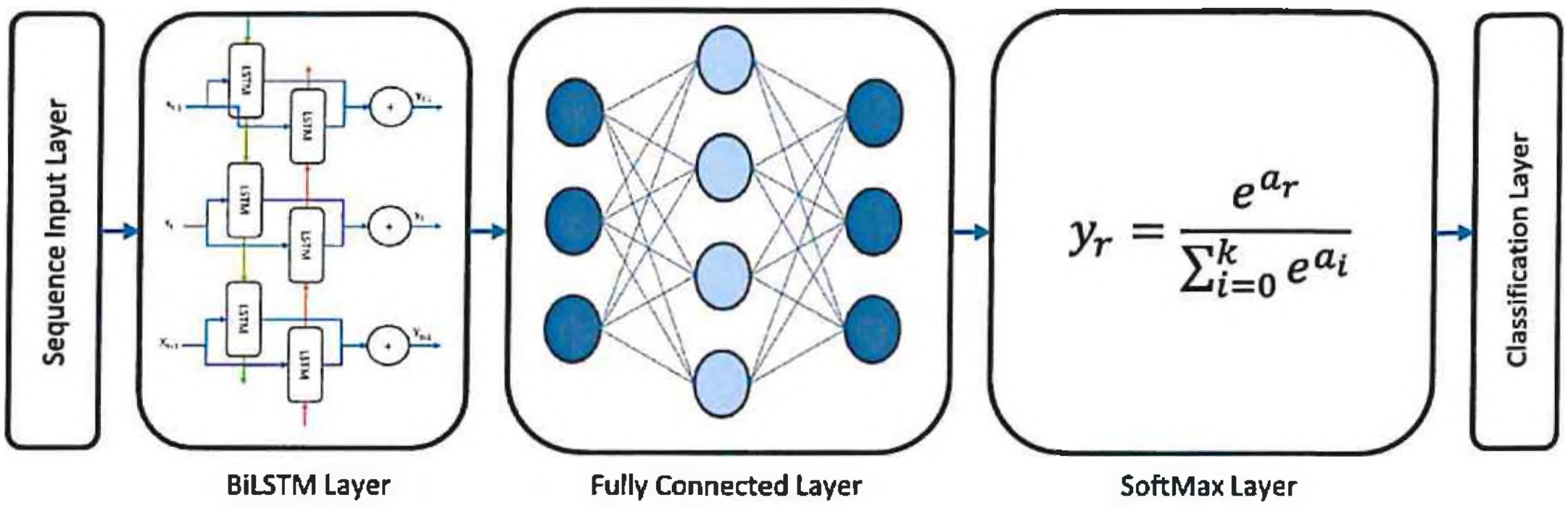

**Fig. 3.** Illustration of the architecture of the bi-LSTM neural network designed for EEG signal classification.

The neural network architecture employed consists of a 61-input-sequence layer. Each input is a row from either matrix **E**(n) or **E**′(n), which were described in Section 3.1.

Although in this article different RNN configurations are compared, the reference configuration is the bi-LSTM (bidirectional-LSTM) layer due to its successful application in other EEG signal applications (Ariza et al., 2023, 2022). This layer is configured with 20 hidden layers, which represents the amount of information that is stored between the various temporal sequences.

Subsequently, a fully-connected layer, that determines the number of outputs ($M$), is added. In this study, different tests have been conducted, with the number of outputs changing for each of them. For the initial test, a five-class classification task is carried out, each corresponding to one of the languages used in this study, resulting in $M = 5$. Following tests involve binary classification tasks, with the number of outputs set to 2. The first binary classification task distinguishes between understanding or not understanding, while the second focuses on differentiating between understanding the native language and understanding a foreign language. Then, a three-class classification task is carried out, with $M$ set to 3. These three classes correspond to three different levels of language comprehension.

Finally, the network architecture is completed with a non-linear softmax layer, and a classification layer that computes the cross-entropy loss for multi-class classification. A visual representation of the network architecture is shown in Fig. 3.

In order to obtain comparative results, the bi-LSTM layer was replaced by two other types of recurrent neural network architectures, which are also commonly used for EEG signal processing: the LSTM layer (Zhang et al., 2022), and the GRU layer (Yang et al., 2023). These layers were configured similarly to the bi-LSTM network, with 20 hidden layers.

## 4. Results

In this section, the results obtained for the different kinds of experiments carried out are presented. Several evaluations for inter-subject scenarios have been performed using the available data from all the subjects. Note that the evaluation of intra-subject scenarios would lead to higher accuracy results on the basis of brain plasticity, but the influence of data variability would be diminished, and direct inter-subject generalization wouldn't be possible. Moreover, conclusions on the performance of features, and classification schemes for the different tasks are better extracted from the inter-subject scenario.

First, a multi-class classification task is performed where subjects listen to sentences in different languages (Section 4.1). As described in Section 2.2, after listening to each sentence, participants assess their level of understanding, with three possible responses: 'understood the meaning', 'understood part of the meaning', or 'did not understand the meaning' of the sentence heard. Using this information, tests have been carried out to investigate whether trials can be classified according to the level of understanding. The results are presented in Section 4.2.

Note that for all the tests, the neural networks maintain the same configuration. The bi-LSTM, LSTM and GRU layers are configured with 20 hidden units, and the network is trained for 5 epochs. Following a conventional procedure to evaluate classification schemes (Theodoridis & Koutroumbas, 2009), in order to carry out the desired evaluations, the dataset used in each test is split into two even subsets: one for training the network and the other for evaluating its performance.

Each evaluation is performed 10 times, and the results presented are the mean values from these repetitions.

Tests were repeated with the three selected neural network architectures: bi-LSTM, LSTM, and GRU layer, under the same conditions.

To represent the results, confusion charts are used, which allow visualizing the performance of algorithms in classification tasks. In order to compare results, the following evaluation metrics are used (Powers, 2011):

- Accuracy is the fraction of correct classifications relative to the total number of classifications.

$$Accuracy = \frac{t_p + t_n}{t_p + t_n + f_p + f_n} \quad (4)$$

  where $t_p$ stands for the number of true positives, $t_n$ for the true negatives, $f_p$ for the false positives, and $f_n$ for the false negatives.
- Recall or true positive rate (TPR) is the probability of positively classifying a sample.

$$Recall = \frac{t_p}{t_p + f_n} \quad (5)$$

- Precision or positive predicted value (PPV) indicates the probability that a sample classified as a given class actually belongs to that class.

$$Precision = \frac{t_p}{t_p + f_p} \quad (6)$$

- F-Score is the harmonic mean of precision and recall, representing the trade-off between these two metrics.

$$F - Score = 2 \cdot \frac{Precision \cdot Recall}{Precision + Recall} \quad (7)$$

### 4.1. Multi-class classification: language

The trials from the conducted experiment (Section 2.2) are classified into 5 different classes, each corresponding to one of the languages used in this study.

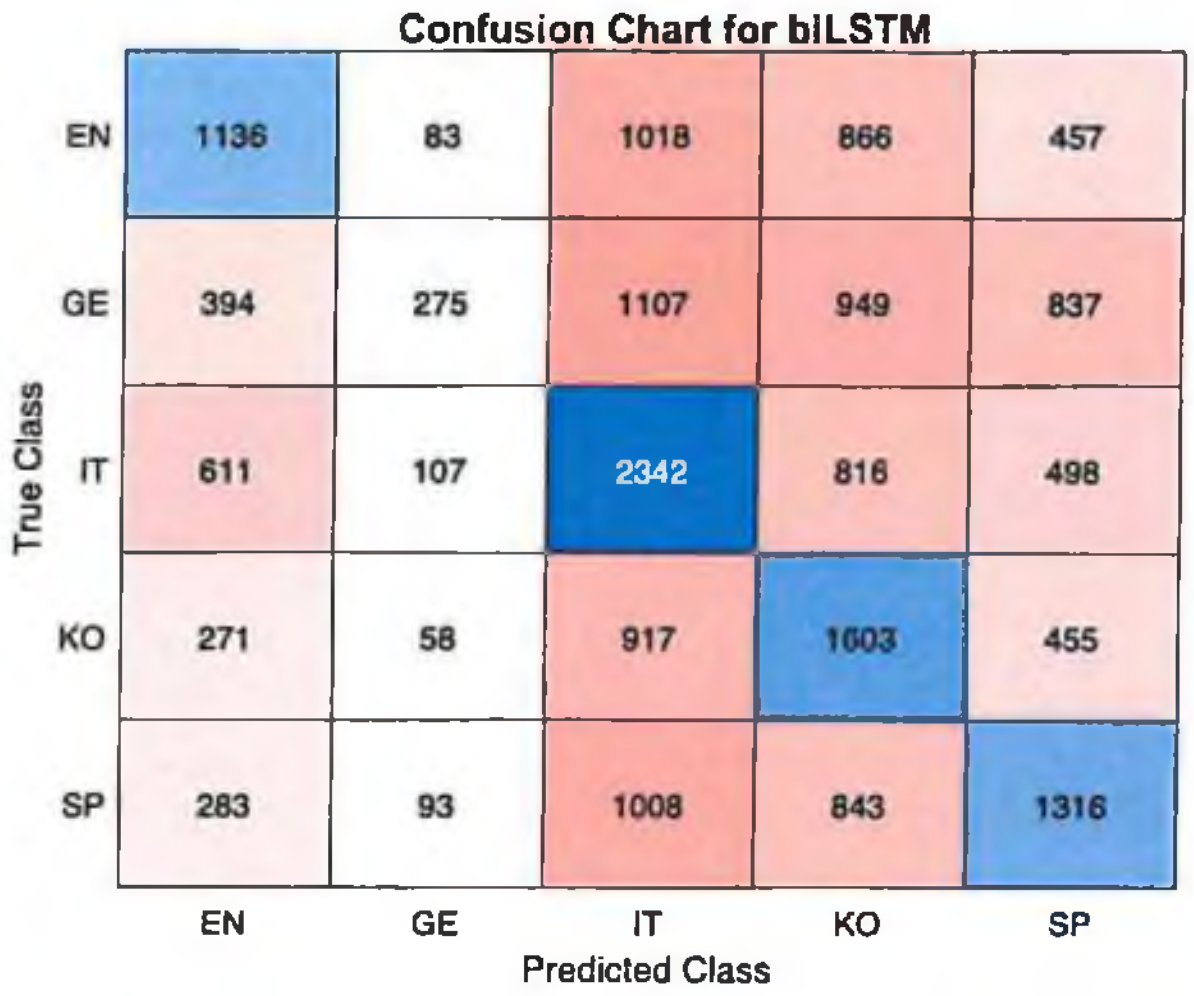

Fig. 4. Confusion chart for multi-class classification using EEG data of subjects listening to sentences in various languages. In this test, the bi-LSTM neural network architecture is used, along with $E'(n)$ obtained from trials of 400 ms.

Two different tests has been carried out in this multi-class classification task scenario. As described in Section 3.1, two matrices have been introduced: $E(n)$, representing the difference of energies between EEG channels of trial $n$, and $E'(n)$, representing the approximation of the derivative of the energy of trial $n$.

Initially, classification is conducted using the matrix $E(n)$. Then, with the aim of observing the temporal evolution of energy in different channels, the matrix $E'(n)$ is used to characterize each trial. These tests have been performed with trials of different durations, 400 ms and 1 s.

In these tests, the objective is to classify EEG signals recorded while subjects were listening to sentences in different languages, using the data obtained from the experiment (see Section 2.2). The trials correspond to 5 different classes, one for each language.

First, $E(n)$ matrix is employed for classifying trials of 400 milliseconds. The accuracy achieved in this case was 30.88 %. However, when $E'(n)$ matrix, incorporating additional temporal evolution information, is used, the accuracy increases to 36.37 %. Fig. 4 shows the confusion chart for this test, conducted with bi-LSTM neural network architecture.

Finally, the tests are replicated using one-second trials. With $E(n)$, an accuracy of 39.87 % was achieved. However, when using $E'(n)$, which incorporates the temporal evolution of the energy, the accuracy rose to 42.91 %.

These tests were also repeated with an LSTM and a GRU layer. The results obtained are shown in Table 4.

Table 4
Accuracy obtained in a five-language classification task with different neuronal network architectures: bi-LSTM, LSTM and GRU.

| Matrix | Trial duration | bi-LSTM | LSTM | GRU |
|---|---|---|---|---|
| $E(n)$ | 400 ms | 30.88 % | 41.24 % | 37.11 % |
| $E(n)$ | 1 s | 39.87 % | 36.62 % | 33.48 % |
| $E'(n)$ | 400 ms | 36.37 % | 52.25 % | 42.51 % |
| $E'(n)$ | 1 s | 42.91 % | 43.42 % | 36.54 % |

Table 5
Results of binary classification task: 'understood' and 'did not understand' using the bi-LSTM neural network architecture.

| Matrix | Duration | Accuracy | F-Score | Recall | Precision |
|---|---|---|---|---|---|
| $E(n)$ | 400 ms | 69.82 % | 71.44 % | 74.20 % | 68.87 % |
| $E(n)$ | 1 s | 67.48 % | 69.93 % | 75.17 % | 65.37 % |
| $E'(n)$ | 400 ms | 73.77 % | 72.23 % | 67.17 % | 78.12 % |
| $E'(n)$ | 1 s | 69.06 % | 71.26 % | 77.51 % | 65.95 % |

Table 6
Results of binary classification task: 'understood' and 'did not understand' using the LSTM neural network architecture.

| Matrix | Duration | Accuracy | F-Score | Recall | Precision |
|---|---|---|---|---|---|
| $E(n)$ | 400 ms | 67.61 % | 70.58 % | 76.81 % | 65.28 % |
| $E(n)$ | 1 s | 64.22 % | 66.01 % | 69.36 % | 62.97 % |
| $E'(n)$ | 400 ms | 73.56 % | 76.28 % | 83.95 % | 69.89 % |
| $E'(n)$ | 1 s | 69.10 % | 71.34 % | 76.33 % | 66.96 % |

Table 7
Results of binary classification task: 'understood' and 'did not understand' using the GRU neural network architecture.

| Matrix | Duration | Accuracy | F-Score | Recall | Precision |
|---|---|---|---|---|---|
| $E(n)$ | 400 ms | 64.69 % | 62.28 % | 57.05 % | 68.57 % |
| $E(n)$ | 1 s | 63.22 % | 66.70 % | 73.53 % | 61.03 % |
| $E'(n)$ | 400 ms | 69.10 % | 68.12 % | 65.87 % | 70.53 % |
| $E'(n)$ | 1 s | 63.86 % | 57.63 % | 48.40 % | 71.21 % |

### 4.2. Tests to evaluate the level of understanding of a language

Using the data obtained from the experiment and the subjects' responses, the trials are labelled according to their level of understanding: trials where the meaning was understood are labelled as 'Y', those where part of the meaning was understood are marked as 'M', and trials where the meaning was not understood are denoted as 'N'. Based on these labels, three different tests have been conducted:

1. Binary classification: distinguishing between whether the subject understood/didn't understand.
2. Binary classification: using the trials where subjects understood the meaning, a binary classification is considered between the categories understanding native language and understanding a foreign one.
3. Multi-class classification: categorizing trials into three categories: understood, understood part of the meaning, or did not understand.

These tests have been performed with trials of 400 ms with an overlap of 50 %, as described in Section 3.1. Additionally, the same tests were repeated using 1-second trials with the same overlap, 50 %. Therefore, each of these tests was carried out under 4 different conditions: using the matrix $E(n)$ and matrix $E'(n)$, with data obtained from both 400 ms and 1 s trials.

As with the previous tests, these results will be compared across different neuronal network architectures: bi-LSTM, LSTM and GRU. The outcomes of these tests are shown in the following subsections.

#### 4.2.1. Binary classification: understood or didn't understand

In this test, language comprehension is assessed by classifying EEG trials into two categories: 'understood' and 'did not understand'. The results for the various scenarios evaluated are presented in Tables 5–7.

The scenario that yielded the highest accuracy occurred when the matrix $E'(n)$ obtained from 400 ms trials was used for classification with the bi-LSTM neural network configuration. Fig. 5 depicts the confusion chart for this specific scenario.

#### 4.2.2. Binary classification: understand native language/understand a foreign language

In this experiment, brain responses when listening to a native language versus a foreign language are taken into consideration by performing a binary classification task into two distinct classes: 'understand native language' (denoted as 'N') and 'understand foreign language' (denoted as 'F'). The results for the different scenarios evaluated using the bi-LSTM, LSTM and GRU neural network architectures are presented in Tables 8–10, respectively.

The scenario that achieved the highest accuracy occurred when the matrix $E'(n)$ obtained from 400 ms trials was used for classification with the bi-LSTM neural network architecture. Fig. 6 depicts the confusion chart for this scenario.

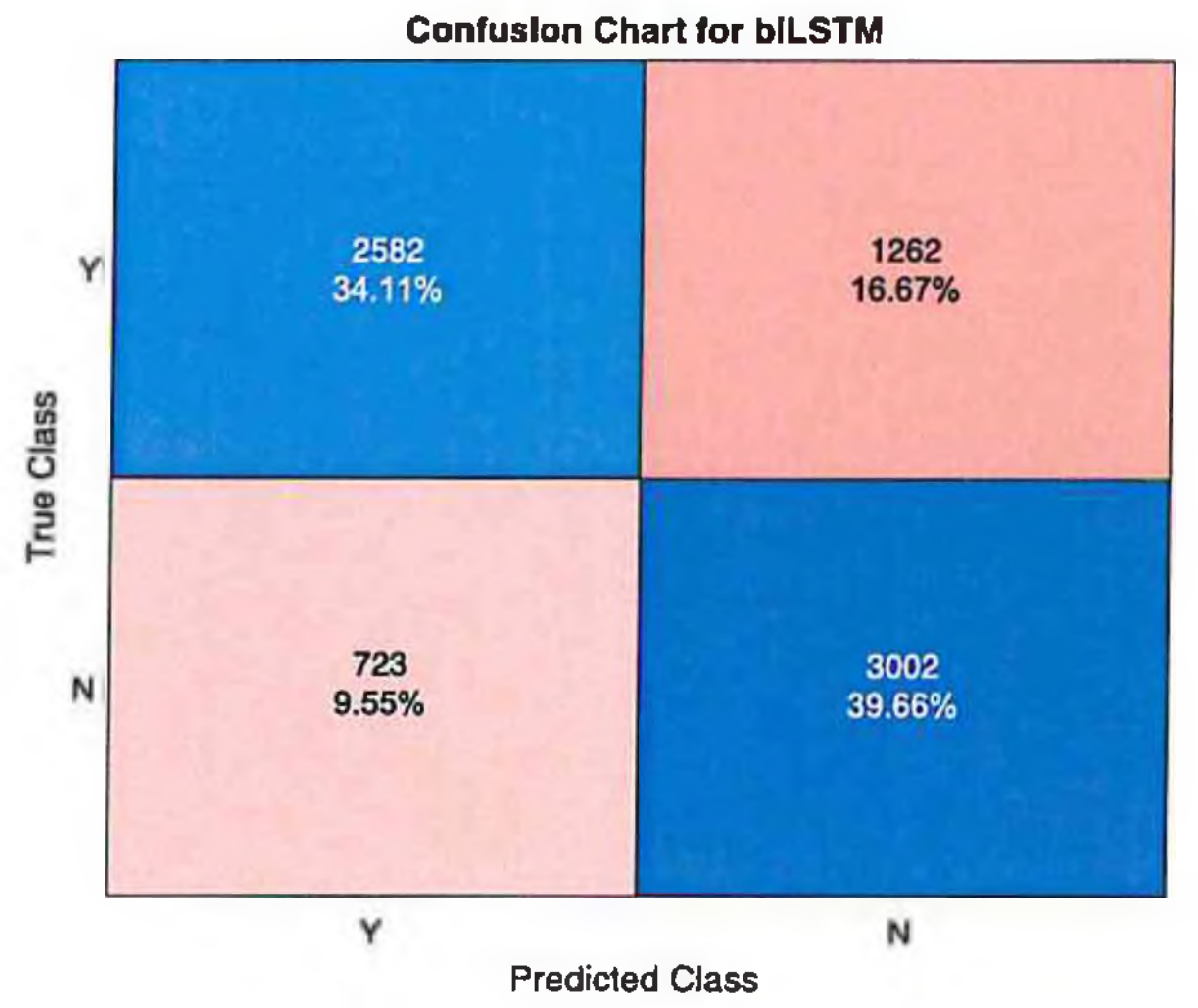

**Fig. 5.** Confusion chart for binary classification using EEG data from subjects listening to sentences in different languages. In this test, $\mathbf{E}'(\mathbf{n})$ obtained from 400 ms trials is used with the bi-LSTM neural network architecture.

**Table 8**
Results of binary classification task: 'understand native language'/'understand foreign language' using the bi-LSTM neural network architecture.

| Matrix | Duration | Accuracy | F-Score | Recall | Precision |
|---|---|---|---|---|---|
| E(n) | 400 ms | 83.42 % | 83.24 % | 82.81 % | 93.67 % |
| E(n) | 1 s | 81.26 % | 80.23 % | 77.29 % | 83.40 % |
| E′(n) | 400 ms | 89.09 % | 89.16 % | 90.15 % | 88.19 % |
| E′(n) | 1 s | 85.43 % | 83.96 % | 77.57 % | 91.49 % |

**Table 9**
Results of binary classification task: 'understand native language'/'understand foreign language' using the LSTM neural network architecture.

| Matrix | Duration | Accuracy | F-Score | Recall | Precision |
|---|---|---|---|---|---|
| E(n) | 400 ms | 81.80 % | 81.73 % | 82.02 % | 81.44 % |
| E(n) | 1 s | 77.78 % | 76.28 % | 72.92 % | 79.97 % |
| E′(n) | 400 ms | 85.80 % | 85.12 % | 84.99 % | 90 % |
| E′(n) | 1 s | 84.53 % | 84.41 % | 84.99 % | 83.84 % |

**Table 10**
Results of binary classification task: 'understand native language'/'understand foreign language' using the GRU neural network architecture.

| Matrix | Duration | Accuracy | F-Score | Recall | Precision |
|---|---|---|---|---|---|
| E(n) | 400 ms | 76.70 % | 79.90 % | 90.75 % | 71.36 % |
| E(n) | 1 s | 73.63 % | 74.43 % | 74.74 % | 74.13 % |
| E′(n) | 400 ms | 83.98 % | 83.42 % | 81.11 % | 85.86 % |
| E′(n) | 1 s | 79.02 % | 80.11 % | 85.37 % | 75.47 % |

### *4.2.3. Multi-class classification: understood/understood part of the meaning/didn't understand*

In this test, language comprehension is assessed by classifying EEG trials into three distinct classes within a multi-class classification task: 'understood'/'understood part of the meaning'/'didn't understand'. The number of trials in which the subject understands part of the meaning is lower than the number of trials for the other classes. To address this imbalance, random under-sampling is applied to ensure an equal number of samples for each class. The results for different scenarios evaluated are exposed in Table 11.

The scenario with the highest accuracy is when the matrix $\mathbf{E}'(\mathbf{n})$ obtained from 400 ms trials, and the bi-LSTM neural network architecture are used to perform the classification. Fig. 7 shows the confusion chart for this scenario.

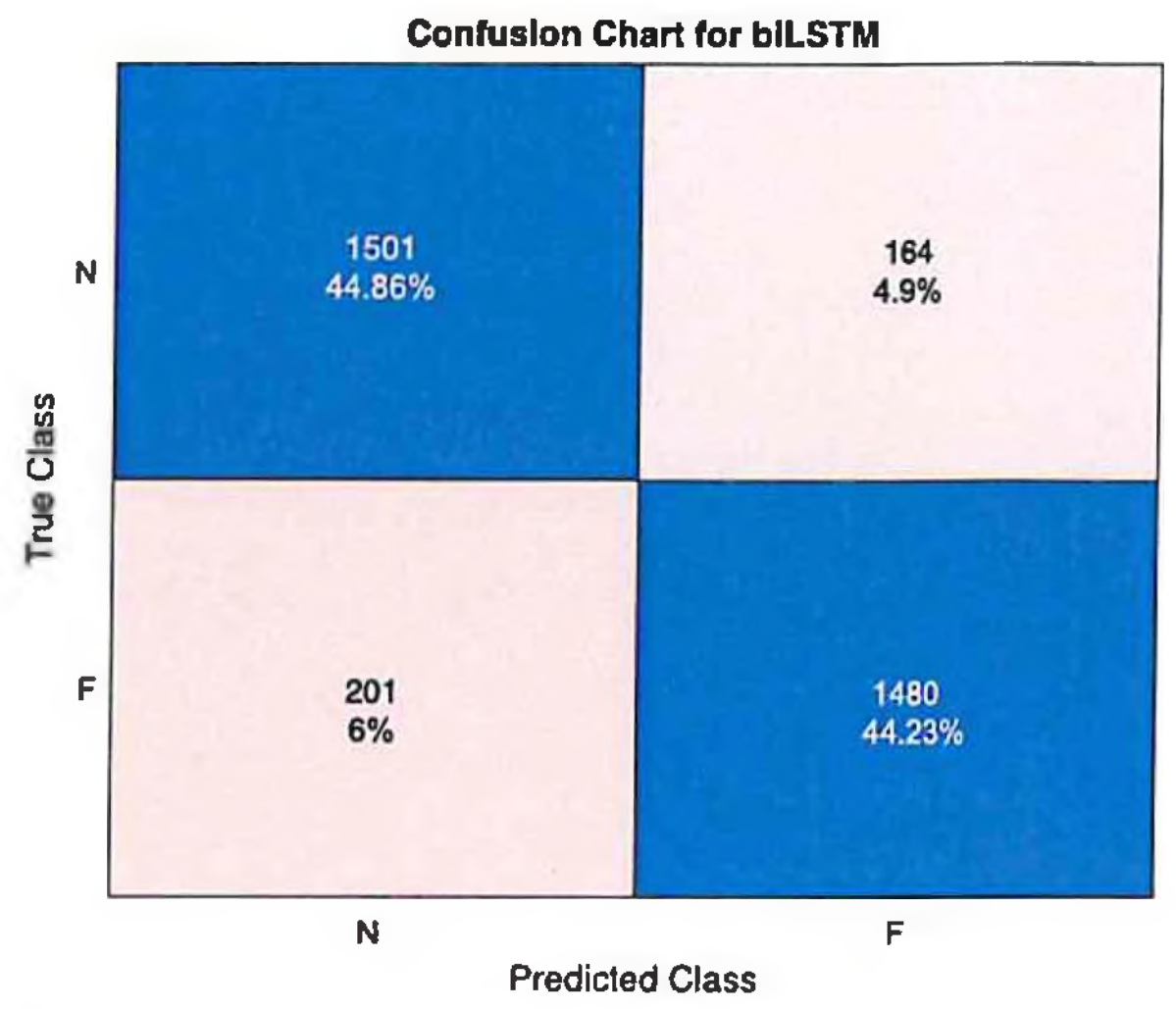

**Fig. 6.** Confusion chart for binary classification using EEG data from subjects listening to sentences in different languages. Test: 'understand native language'/'understand foreign language'. In this experiment, $\mathbf{E}'(\mathbf{n})$ obtained from 400 ms trials, and the bi-LSTM neural network architecture are used for classification.

**Table 11**
Accuracy obtained in multi-class classification: understood/understood part of the meaning/didn't understand, with different neural network architectures: bi-LSTM, LSTM and GRU.

| Matrix | Duration | bi-LSTM | LSTM | GRU |
|---|---|---|---|---|
| E(n) | 400 ms | 74.51 % | 72.02 % | 69.99 % |
| E(n) | 1 s | 71.52 % | 68.50 % | 66.92 % |
| E′(n) | 400 ms | 81.36 % | 77.47 % | 76.57 % |
| E′(n) | 1 s | 76.43 % | 71.53 % | 68.69 % |

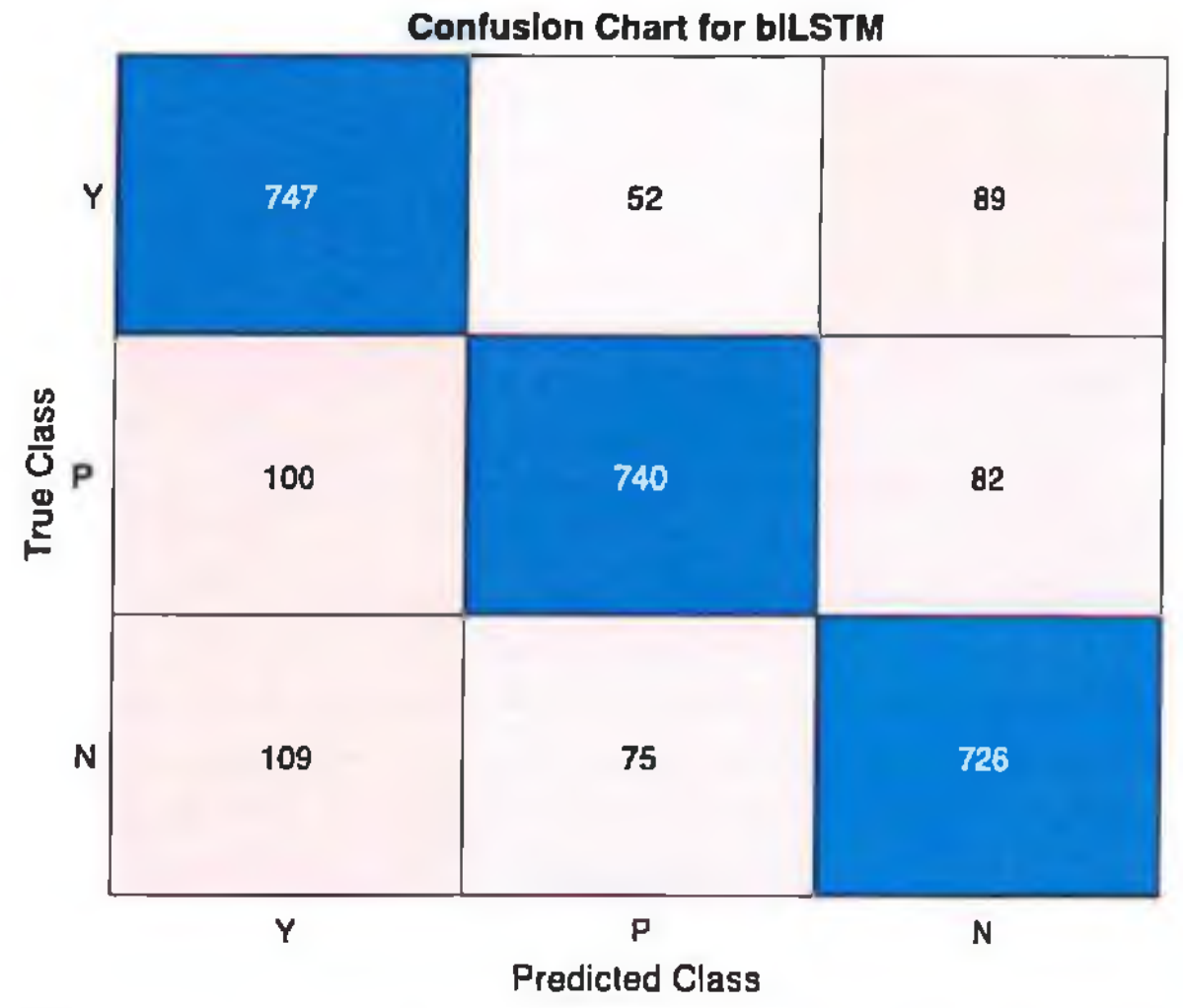

**Fig. 7.** Confusion chart for multi-class classification using EEG data from subjects listening to sentences in different languages. Test: 'understood'/'understood part of the meaning'/'didn't understand'. In this test, $\mathbf{E}'(\mathbf{n})$ obtained from 400 ms trials, and the bi-LSTM neural network architecture are used for classification.

### 4.3. Discussion

The aim of this research is to investigate whether the human brain responds differently to different languages, and whether the level of understanding can be assessed by using EEG signals. To achieve these objectives, we conducted performed two types of classification tasks: binary and multi-class. In these tests, trials of different durations, 400 ms and 1 s, were used, along with the analysis of the evolution of energy differences between consecutive trials. Additionally, different neural network configurations were evaluated: bi-LSTM, LSTM and GRU.

In the multi-class scenario, we conducted tests to observe how our brain responds when listening to sentences in different languages. EEG signals are then used to categorize trials according to the language the subjects were exposed to.

In order to carry out these experiments, a database was created with data from six subjects listening to sentences in different languages and giving feedback on whether or not they understood the meaning of each sentence. The dataset consists of 18653 trials of 400 ms and 8463 trials of 1 s, providing a representative sample for analysis. It should be noted that all the tests performed in this study are inter-subject tests, which introduces variability into the data.

In these language listening tests, the accuracy achieved with the bi-LSTM neural network architecture was 36.37 %. These results suggest that the brain reacts differently when listening to different languages, as each language has its own rhythm and musicality. It should be acknowledged that factors such as the level of understanding of a language, and its similarity to the subjects' native language play a role in this problem.

Regarding the experiments described in Section 4.2 on the level of comprehension of listened language, several conclusions can be drawn. First, accuracy is higher when the trial duration is shorter; specifically, in our tests, when 400 ms trials are used to perform the classification task the accuracy increases to 89,09 %. Nevertheless, it is important to note that the number of trials is smaller when the trial duration is 1 s. Second, the best results are achieved when using the average of the derivative of the energy of trials ($\mathbf{E}'(n)$) for classification across different tests. This suggests that predictions are more accurate when classifying the evolution of energy information from both the current and previous samples.

The results of the first test (Section 4.2.1) and the third test (Section 4.2.3) reveal that that different levels of language comprehension elicit distinct brain responses. Moreover, at sight of the best accuracy attained in the experiments when classifying trials into understand native language/understand foreign language (Section 4.2.2) (89,09 % with 400 ms trials, $\mathbf{E}'(n)$ and bi-LSTM neural network configuration), it can be concluded that the human brain reacts differently when listening to sentences in their native language as opposed to a foreign language, even when the meaning is understood.

Levari and Snedeker (2024) explain that when adults listen to a story, they use the context to predict the next words. This aligns with the results obtained in all the tests conducted, which show better performance when the average of the derivative of the energy of the trials is used, i.e. when context is included rather than just the energy of a particular trial.

The accuracy obtained in the native language vs. foreign language classification is comparable to that reported by Sakthi et al. (2019) despite differences in testing protocols and EEG signal processing. The tests and EEG signal processing in Sakthi et al. (2019) differ from those presented in this article (e.g., five languages are used, some of which share similarities, such as Spanish and Italian, whereas only English narratives with a disturbing tone are used in their study, and the subjects' mother tongues are either English or Chinese). This further reinforces the idea that the brain processes the mother tongue differently compared to a foreign language.

Comparing the results obtained with the different architectures, it is observed that in all cases the results are higher when using a LSTM architecture, with the bi-LSTM yielding the best performance. Therefore, we can conclude that this architecture is more suitable than the GRU architecture for this type of classification. This contrasts with the findings in references Zhang et al. (2022) and Yang et al. (2023), where the GRU network achieved better accuracy for motion prediction in beating heart analysis.

Overall, the best accuracy for both binary and multi-class classification is achieved using the bi-LSTM neural network architecture along with the $\mathbf{E}'(n)$ matrix. Additionally, a high F-score value was obtained across the different architectures, indicating that the proposed method exhibits good classification performance for the task at hand. Therefore, it can be confirmed that recurrent neural networks are suitable for the analysis of EEG signals, using specific measures of relationships between energy in the different EEG channels for brain activity characterization. Also, it must be noted that the results improved when temporal progression was explicitly considered in the study by analysing the evolution of the energy difference arrays between consecutive samples.

Regarding the computational complexity of this network, it can be measured by the number of parameters involved. This number, in our bi-LSTM network (see Fig. 3) with 61 inputs, 2 outputs and 20 hidden units, is limited to 13240 for our Matlab R2024a implementation. Also, the computational load of the proposed processing scheme for classification, once the networks are trained, was found to be small, achieving a classification speed much faster than the acquisition rate, even in a low-performance computer (e.g. a laptop with an AMD Ryzen 7 5000 series processor running Windows 11 and Matlab R2024a).

## 5. Conclusion

In this paper, we explored brain responses to different languages through the analysis of EEG signals. To achieve this, we designed an experiment where subjects listened to phrases in various languages. To characterize the EEG signals, we constructed a matrix representing the energy differences between the different channels, and then used bi-LSTM, LSTM and GRU neural networks to classify the samples, with the bi-LSTM network achieving the best results. Additionally, explicit temporal evolution of energy was considered by using an approximation of the derivative, which lead to improved performance.

In order to analyse the brain's response, a database of EEG signals has been created (Ariza et al., accessed 10 March 2025). This database consists of the EEG signals from six subjects participating in the experiment described in Section 2.2, where they listened to sentences in five different languages. In addition, the database includes feedback from the subjects on whether they understood each sentence, or not.

This approach was used to conduct multi-class classification tests. In the first test, EEG signals were classified into 5 distinct groups, one for each language, achieving an accuracy over 36 %. In a second test, which evaluated whether the subjects understood or did not understand the sentences, the accuracy reached 73.77 %. In the test distinguishing between understanding a native language or a foreign one, the accuracy surpassed 89 %.

These results confirm that the human brain reacts differently when listening to sentences in a native language compared to a foreign language, even when the meaning is understood. These differences can be observed by means of the analysis the corresponding EEG signals. Specifically, energy relationships across EEG channels, combined with the bi-LSTM neural network framework for classification, lead to these conclusions.

The best results in this study pertain to the classification between understanding the native language or a foreign language. This is consistent with the findings in reference Sakthi et al. (2019). When comparing the results obtained with the neural network architectures considered, bi-LSTM, LSTM and GRU, it is observed that the highest accuracy is achieved using the bi-LSTM architecture.

Also, results are in concordance to Levari and Snedeker (2024), who suggested that the cognitive process of understanding a story is aided by context to predict the following words, the accuracy obtained improves in all cases using the average of the derivative of the energy of the trials.

Note that all the experiments conducted in this studio are inter-subject experiments, which introduce more variability in the data compared to intra-subject scenarios.

The results obtained in this work are based on data gathered from experiments carried out by six subjects, most of whom are Spanish, who listened to sentences in five different languages. The number of subjects and the specificity of their knowledge of the languages constitute the main limitation of our research. It could be extended and strengthened by expanding the database to include more subjects whose native language is other than Spanish or by incorporating languages from other countries.

The present study serves as the foundation for the design of new experiments, such as one where subjects begin learning a language they do not know. This would allow for the analysis of the evolution of language comprehension levels alongside the corresponding EEG signals. Such an approach could provide insights into how the learning curve of a language is reflected in EEG signals and how to analyse them. If such experiment were performed on different subjects employing diverse learning methods, it could potentially lead to improvements in learning methodologies. Also, this could serve as a foundational point for research on the understanding of auditory stimuli in patients in a state of minimal consciousness.

The next step for further progress in this field is to enlarge the database and identify the specific areas of the brain responsible for understanding sentences. This would help minimize the number of EEG channels required, facilitating the development and application for both medical and educational tools.

## CRediT author statement

Isaac Ariza: Conceptualization, Methodology, Software, Validation, Formal analysis, Investigation, Data curation, Writing, visualization. Lorenzo J. Tardón: Conceptualization, Methodology, Software, Investigation, Formal analysis, Writing, Visualization, Data curation, Project administration. Ana M. Barbancho: Conceptualization, Methodology, Software, Investigation, Validation, Formal analysis, Data curation, Writing, Visualization. Isabel Barbancho: Conceptualization, Methodology, Software, Investigation, Formal analysis, Validation, Writing, Visualization, Data curation, Resources, Project administration, Funding acquisition.

## Data availability

A link to the data is supplied.

## Declaration of competing interest

The authors declare that they have no known competing financial interests or personal relationships that could have appeared to influence the work reported in this paper.

## Acknowledgments

This publication is part of project PID2021-123207NB-I00, funded by MCIN/AEI/10.13039/501100011033/FEDER, UE. This work was done at Universidad de Málaga, Research Group TIC208, Campus de Excelencia Internacional Andalucía Tech. Funding for open access charge: Universidad de Málaga/CBUA.

## Appendix A. Sentences used in the listened-language experiment

**Table A.1**
Sentences in English used in the experiment (2.2).

| Transcription | Duration |
|---|---|
| He fell asleep, although the TV was on. | 0:05 |
| It was already late. Nevertheless, he stayed a while. | 0:06 |
| Despite having no license, he drives the car. | 0:05 |
| I don't know if he'll come back. | 0:03 |
| Maybe he won't come back. | 0:03 |
| I wonder if he thinks about me. | 0:04 |
| What is the name of the capital city of Switzerland? | 0:05 |
| What are the doctor's consultation times? | 0:04 |
| He is going to stay either with us or in the hotel. | 0:04 |
| I become tired as soon as I have to study. | 0:03 |
| She reads the newspaper instead of cooking. | 0:03 |
| I overslept. Otherwise, I'd have been on time. | 0:04 |
| During the holidays, the children were allowed to remain outside late. | 0:05 |
| Were you allowed to drink beer in the hospital? | 0:03 |
| You speak so softly. Don't speak so softly. | 0:06 |
| You are so lazy. Don't be so lazy. | 0:05 |
| I want to deposit money in my account. | 0:03 |
| No, I don't understand them so well. | 0:03 |
| She has lived in Madrid as well as in London. | 0:03 |
| Are tickets for the theatre still available? | 0:03 |
| He was disloyal, but she was loyal. | 0:05 |
| He had no money, only debts. | 0:05 |
| She listens to music while she does her work. | 0:04 |
| I can't smell anything when I have a cold. | 0:05 |
| After he had lost his job, he went to America. | 0:03 |
| I'm not drinking it because I have to drive. | 0:04 |
| Is the exhibition open on Tuesdays? | 0:04 |
| I'm not eating it because I must lose weight. | 0:05 |
| I want to go to the bookstore to buy a book. | 0:05 |
| Drive until you reach the third traffic light. | 0:03 |

**Table A.2**
Sentences in German used in the experiment (2.2).

| Transcription | Duration |
|---|---|
| Ich arbeite bis 13.30 Uhr und bin um 14:00 Uhr zu Hause | 0:04 |
| Es gibt einen Kurs montags, mittwochs und freitags von 14:30 bis 18:00 Uhr. | 0:05 |
| Viele Touristen reisen nach Bern, und auch viele Besucher aus der Region. | 0:05 |
| Ich weiß nicht, ob er mich liebt. | 0:03 |
| Ich frage mich, ob er an mich denkt. | 0:04 |
| Es ärgert mich, dass du schnarchst. | 0:04 |
| Das Wetter wird vielleicht morgen besser. | 0:04 |
| Ich finde, dass er sogar sehr gut aussieht. | 0:05 |
| Es ist gut möglich, dass er eine Freundin hat. | 0:05 |
| Ich warte, bis meine Haare trocken sind. | 0:03 |
| Ich warte, bis der Film zu Ende ist. | 0:04 |
| Seit wann arbeitet sie nicht mehr? | 0:03 |
| Sie arbeitet nicht mehr, seitdem sie geheiratet hat. | 0:04 |
| Seitdem sie sich kennen, sind sie glücklich. | 0:04 |
| Seitdem sie Kinder haben, gehen sie selten aus. | 0:04 |
| Hast du einen Flaschenöffner? | 0:03 |
| Der Fernseher war an; trotzdem ist er eingeschlafen. | 0:04 |
| Das Hotel war zwar gemütlich, aber zu teuer. | 0:04 |
| Er nimmt entweder den Bus oder den Zug. | 0:04 |
| Sie spricht sowohl Spanisch als auch Englisch. | 0:04 |
| Er ist nicht nur dumm, sondern auch faul. | 0:03 |
| Sie ist nicht nur hübsch, sondern auch intelligent. | 0:04 |
| Ich kann weder Klavier noch Gitarre spielen. | 0:04 |
| Mein Sohn wollte nicht mit der Puppe spielen. | 0:04 |
| Meine Kinder wollten keinen Spaziergang machen. | 0:04 |
| Durftest du den Hund ins Hotel mitnehmen? | 0:04 |
| In den Ferien durften die Kinder lange draußen bleiben. | 0:04 |
| Ich dachte, du wolltest deine Frau anrufen. | 0:04 |
| Welche Kreditkarten kann man benutzen? | 0:04 |
| Ich esse sie nicht, weil ich abnehmen muss. | 0:04 |

**Table A.3**
Sentences in Italian used in the experiment (2.2).

| Transcription | Duration |
|---|---|
| Milano è una grande città nel nord Italia. | 0:05 |
| Gli abitanti di Milano si chiamano milanesi. | 0:04 |
| La prima stagione dell'anno è la primavera. | 0:05 |
| Io sono a casa con mia moglie e siamo raffreddati. | 0:05 |
| Questa sedie è di legno e quel tavolo è di plastica. | 0:05 |
| Queste scarpe sono marroni e quei pantaloni sono blu. | 0:04 |
| Mi piace questa canzone e odio quel film. | 0:04 |
| Il cane si ha spaventato per i petardi. | 0:04 |
| Sono di Cagliari, la città più importante della Sardegna. | 0:04 |
| Sono le due di notte e il cane dei vicini non smette di abbaiare. | 0:05 |
| Non sono mai stato a Pamplona, ma vorrei andarci per vedere la corsa dei tori. | 0:05 |
| Hanno aperto un nuovo centro commerciale proprio proprio dietro alla piazza principale. | 0:05 |
| In Piazza Garibaldi ci sono i Carabinieri, l'ufficio postale e la biblioteca. | 0:06 |
| Questa mattina sono andato in Piazza Garibaldi a presentare di documenti in municipio. | 0:05 |
| Prima di ritornare a casa sono passato per Piazza Marco Polo e mi sono fermato a l'ufficio postale per spedire una lettera. | 0:08 |
| Quando era bambina ricordo che i miei genitori mi portavano in vacanza al mare. | 0:05 |
| A me piaceva tuffarmi nel agua che se era fredda, a mia sorella invece il mare faceva un può di paura. | 0:06 |
| Dopo una mezz'ora ci asciughiamo e torniamo in albergo per il pranzo. | 0:05 |
| Alle quatre tornavamo in spiaggia e ci tuffavamo subito in mare per il bagno di pomeriggio. | 0:05 |
| Dopo aver cenato, ricordo che andavamo tutti insieme a fare una passeggiata per la città. | 0:05 |
| Alle nove di solito arrivano anche la seggretaria ed il direttore. | 0:05 |
| Alle due preparo un report su la produzione dell ultimo mese. | 0:05 |
| Di primo abbiamo la lasagna al ragu, penne al pomodoro e gnocchi al pesto. | 0:05 |
| Non ho capito qual è il nome della stazione della quale ha partito il suo treno. | 0:05 |
| Fare il enfermiere non è il lavoro che io sempre ho sognato di fare. | 0:04 |
| Gli studenti con i quali vai a lezioni sono tutti più giovani di te. | 0:04 |
| L'aeroporto verso cui si dirige l'aereo del presidente è il più importante del paese. | 0:06 |
| Il film del quale mi hai parlato non fa per me; preferisco altre generi. | 0:05 |
| Ho comprato un nuovo telecomando universale che funziona con tutte le televisioni che abbiamo in casa. | 0:06 |
| Giovanni dice che la sua ricetta preferita è quella che gli ha insegnato Maria. | 0:05 |

**Table A.4**
Sentences in Korean used in the experiment (2.2).

| Transcription | Duration |
|---|---|
| Naneun han-gugeo-reur chogeum har su iseumnida | 0:03 |
| Yeogiseo inteones-eur iyong-har su iseumni-kka? | 0:03 |
| Kan-jang-gong-jang kong-jang jang-eun kang kong jang-jang-igo. | 0:04 |
| Tashi malsseumae juseyo. | 0:03 |
| Choe kajogi yogui issoyeo. | 0:04 |
| Onje chorel jan bon panmunaseyo. | 0:03 |
| Urinen jakyoe issoyo. | 0:03 |
| Urinen saramdelgua marago shipoyo. | 0:04 |
| Urinen panguasari piriojeyo. | 0:04 |
| Urinen yorume sanchek kanengosel choajeyo. | 0:04 |
| Kyorenen nuni ogona piga huayo. | 0:04 |
| Chonen clashic umaguel choajeyo. | 0:04 |
| Chonen chiguem i chegl ilgo issoyeo. | 0:04 |
| Chaueguiga innenbanguel uanjeyo. | 0:03 |
| Myoshie sonyok shiksarel choyo? | 0:03 |
| Koguirel annon kosel chuseyo. | 0:03 |
| Senkrim onyen aiskrimel chuseyo. | 0:04 |
| Kichaga onje Vienae totchakeo? | 0:04 |
| Yogui choen shiktangui odi issoyeo? | 0:04 |
| Odissoe kochel salso issoyeo? | 0:03 |
| Chisangui irioil mada yoroyo? | 0:03 |
| Onel choniogue yongjaguaneso mol sanguonjeyo? | 0:04 |
| Toguil timi yonguk timgua yonguihago issoyeo. | 0:04 |
| Yoguie chogumdo momurel koyeyo? | 0:04 |
| Do nuga kopirel mashigo shipoyo? | 0:03 |
| Cho yochaie pomonimi nuguseyo? | 0:03 |
| Kenyoi punmunyim chibe ottoke kayo? | 0:04 |
| Ke chibun ke kireil kette issoyeo. | 0:04 |
| Chochennaren uorioirieyo. | 0:03 |
| Ilchoirenen chiriri issoyeo. | 0:03 |

**Table A.5**
Sentences in Spanish used in the experiment (2.2).

| Transcription | Duration |
|---|---|
| Yo ayudo a mi padre en la cocina. | 0:04 |
| Mi hermana estudia en su habitación. | 0:04 |
| Mi madre lava el coche verde. | 0:04 |
| Mi abuela pasea todos los días. | 0:04 |
| Los perros juegan en el jardín. | 0:04 |
| El cerdo come bellotas. | 0:04 |
| El conejo es veloz. | 0:04 |
| Me levanto a las 8:00 de la mañana. | 0:04 |
| El próximo 4 de octubre José Sáez publica su tercer disco. | 0:04 |
| El Congreso español prohibió fumar en sus instalaciones, anticipándose a una ley del Gobierno que prohíbe el cigarrillo en los sitios de trabajo de toda España. | 0:08 |
| No sé lo que tiene todavía, pero le duele mucho la cabeza y la garganta. | 0:06 |
| Tome estas pastillas tres veces al día después de comer algo. | 0:04 |
| Mi compañero tiene una niña muy bonita. | 0:04 |
| El menú de 14 euros consiste en sopa de pescado, lenguado y tarta de manzana. | 0:07 |
| Como ve, el piso es pequeño pero está muy limpio. | 0:03 |
| Desde las 2 hasta las 5 las tiendas están cerradas. | 0:05 |
| En cuanto llegues, quiero que prepares la merienda para Luisito y su amigo Jaime. | 0:04 |
| Después de merendar, no quiero que vean la tele; diles que hagan los deberes primero. | 0:05 |
| Hay unos filetes de pescado en la nevera, pero no creo que haya ninguna verdura fresca. | 0:05 |
| En febrero se celebra el Carnaval en casi todas las ciudades de España. | 0:04 |
| Durante la Feria de Abril de Sevilla, los sevillanos llevan los trajes típicos de Andalucía, bailan sevillanas y flamenco y montan a caballo. | 0:08 |
| En agosto y septiembre hay muchas fiestas y verbenas locales por todo el país. | 0:05 |
| A mí no me gusta nada la música clásica. | 0:04 |
| A mí me gustan mucho las películas de los hermanos Cohen. | 0:04 |
| Dentro de la ciudad, el metro es el medio de transporte más rápido y popular. | 0:05 |
| Funciona desde la 6.00 de la mañana hasta la 1.30 de la madrugada. | 0:05 |
| Además, Madrid tiene 165 líneas de autobuses que funcionan entre las 6.00 y las 24.00 horas. | 0:07 |
| Esos girasoles y esos geranios son gigantes. | 0:04 |
| La gente de Gibraltar habla inglés y español. | 0:04 |
| ¿Podría hablar con el señor López, por favor? | 0:03 |